\documentclass[12pt]{article}

\usepackage{epsfig}

\usepackage{amssymb}

\def\be{\begin{equation}}
\def\ee{\end{equation}}
\def\ba{\begin{array}{c}}
\def\ea{\end{array}}

\def\ben{$$}
\def\een{$$}

\newcommand{\bea}{\begin{eqnarray}}
\newcommand{\eea}{\end{eqnarray}}

\newcommand{\bbr}{\br\!\br}
\newcommand{\kkt}{\kt\!\kt}
\newcommand{\pbr}{\prec\!}
\newcommand{\pkt}{\!\!\succ\,\,}
\newcommand{\kt}{\rangle}
\newcommand{\br}{\langle}

\begin{document}

\begin{center}

{\Large \bf

Wheeler-DeWitt equation
and the applicability of
crypto-Hermitian
interaction representation in quantum cosmology

  }

\vspace{9mm}

{Miloslav Znojil}

\vspace{9mm}

\vspace{0.2cm}

Department of Physics, Faculty of Science, University of Hradec
Kr\'{a}lov\'{e},

Rokitansk\'{e}ho 62, 50003 Hradec Kr\'{a}lov\'{e},
 Czech Republic

\vspace{0.2cm}

 and

\vspace{0.2cm}

The Czech Academy of Sciences, Nuclear Physics Institute,

 Hlavn\'{\i} 130,
250 68 \v{R}e\v{z}, Czech Republic

\vspace{0.2cm}

%
%
%
%

{e-mail: znojil@ujf.cas.cz}


\end{center}

\newpage

\section*{Abstract}

In the broader methodical framework of quantization of gravity
the crypto-Hermitian (a.k.a.
non-Hermitian) version of the Dirac's
interaction picture
is considered. The formalism is briefly outlined and
shown well suited for an innovative treatment
of certain
cosmological models.
In particular, it is demonstrated that
the
Wheeler-DeWitt equation
could be a promising candidate for the description of
evolution of quantized Universe near its initial
Big Bang singularity.

\section*{Keywords}
.

quantum gravity and the problem of Big Bang;

hiddenly Hermitian formulations of
quantum mechanics;

stationary Wheeler-DeWitt system;

physical Hilbert space metric;

non-stationary Wheeler-DeWitt system;

\newpage

\section{Introduction \label{oh} }

The concept of the wave function $\psi$ of the Universe (introduced,
55 years ago, as a solution of the Einstein-Schr\"{o}dinger {\it
alias\,} Wheeler-DeWitt (WDW) equation \cite{WDW,WDWb}) is contradictory.
On positive side this concept played a key role during the
development of the canonical quantization of gravity
\cite{Thiemann}. These efforts climaxed in the recent comparatively
satisfactory and constructive formulation of the so called loop
quantum gravity (LQG, \cite{Rovellib,Rovelli,Rovellic}). At the same time,
Mostafazadeh pointed out, in his review of the recent progress in
quantum theory \cite{ali}, that the solutions $\psi$ themselves
remain ``void of a physical meaning'' without ``finding an
appropriate inner product on the space of solutions of the WDW
equation'' (see p. 1291 in review \cite{ali}).
In {\it loc. cit.} Mostafazadeh also emphasized that ``the lack of a
satisfactory solution to this problem has been one of the major
obstacles in transforming canonical quantum gravity and quantum
cosmology into genuine physical theories''. Precisely this obstacle
is to be addressed and discussed in what follows.

In the cited review we can further read that ``in \ldots quantum
cosmology \ldots the relevant field equations \ldots are second
order differential equations in a ``time'' variable \ldots [which]
have the \ldots general form
 \be
 \frac{d^2}{dt^2} \psi(t)+D(t)\,
 \psi(t)=0
 \label{WDWE}
 \ee
where $t$ denotes a dimensionless time variable, $\psi: \mathbb{R} \to {\cal L}$
is a function taking values in some separable Hilbert space
${\cal L}$ , and $D : {\cal L}\to {\cal L}$ is a positive-definite operator
that may depend on $t$''. Treating the latter variable
as ``a fictitious evolution parameter in quantum cosmology''
(see p. 1292 in \cite{ali}),  the same author later adds that
``the cases in which $D$ is $t-$dependent (that arises in quantum cosmological models)
require a more careful
examination''.
In this sense we are prepared to discuss some of the open questions and subtleties of
the theory.

In {\it loc. cit.},
Mostafazadeh
redirected interested readers to his earlier study \cite{aliWDW}.
In a series of our subsequent unpublished comments on this topic \cite{which}
(which were later finalized and summarized in papers \cite{timedep,SIGMA})
we showed that an appropriate ``dealing with these cases'' is, {\em simultaneously}, less
complicated and more
complicated than it seems. Less
complicated in the sense that
some of the technical obstacles have later been found surmountable,
and more complicated because it appeared necessary to
amend the overall quantum-theoretical framework and to
replace the non-Hermitian Schr\"{o}dinger-picture (NSP)
interpretation of
the evolution of $\psi$
(as presented, basically, in \cite{ali} or \cite{aliWDW})
by the more involved formalism called
interaction picture (IP) (a.k.a. Dirac's
representation; see a comprehensive review
of its non-Hermitian form (NIP) in \cite{NIP}).

In what follows we intend to outline the implementation of the NIP
approach in the WDW case.
The key purpose of our paper is to provide an explicit
explanation of the connection between several
challenging and
open physical questions
(a typical one
concerns the quantum Big Bang problem as
formulated in section \ref{hown})
and the most recent progress
in the hiddenly unitary version of quantum mechanics
(the basic features of this theoretical innovation are
reviewed).
Our main message (viz, the detailed description
of the theory and of its application to the WDW equation)
will finally be outlined in
section \ref{dodada} (devoted to a
specific schematic toy-model of the
quantum geometry of the Universe),
in section \ref{seci2} (on the full-fledged NIP formalism),
and in section
\ref{seci3} (in which the mechanism of transition to
the Big Bang singularity will be given its ultimate
model-independent construction-recipe form).
Our results will be
discussed in section \ref{discussion}
and summarized in section~\ref{seci5}.


\section{Challenge: Quantum Big Bang problem \label{hown} }


At present it is widely believed that
up to the ``youngest age'' of the Universe (i.e.,
for times $t>t_1$
with $t_1\approx 10^{32}$ seconds) the evolution (i.e.,
a slow expansion) of the Universe is more or less
safely controlled by the classical theoretical cosmology.
In
contrast, in the
interval of times $(t_0,t_1)$ (where $t_0=0$ denotes a hypothetical
time of Big Bang)
we still miss a fully consistent and rigorous quantum
theory behind the early history of the Universe \cite{Thiemann}.

\subsection{Could the degeneracy survive quantization? Yes, it could.}

In our present study we felt strongly motivated by the deep
relevance
of the understanding of the
evolution of the Universe near its Big Bang origin,
i.e., in a genuine quantum dynamical regime.
In this regime the theoretically
most ambitious LQG formalism still
seems to lead to at least some contradictory results.
In one of the LQG predictions \cite{AshteBi},
for example, the Big Bang singularity
(compatible with the classical Einstein's theory of gravity)
has been found to be smeared out by the quantization.
In
the series of papers \cite{loopash,loopashb,loopashc,piech,loopashd,loopashe}
or in section Nr. 8 in \cite{Rovelli}),
for example, it is claimed that
the
Big Bang singularity of
classical theory
must necessarily be
replaced by a regularized ``Big Bounce''
mechanism.
In contrast,
more recently, Wang with Stankiewicz \cite{nobounce}
came with an opposite conclusion claiming that
within the scale-invariant LQG framework
``the quantized Big Bang is {\em not\,} replaced by a Big Bounce''.


\begin{figure}[h]                    
\begin{center}                         
\epsfig{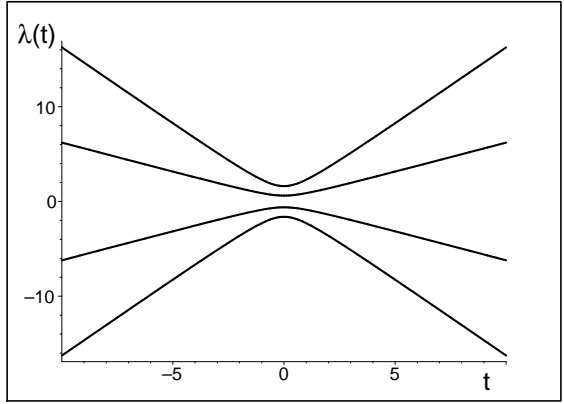}
\end{center}    
\vspace{2mm} \caption{Eigenvalues
of matrix (\ref{tyome})
(avoided crossing phenomenon).
 \label{uba2}
 }
\end{figure}

At the first sight, the latter claim looks suspicious.
In Rovelli's words, the quantization-related
``absence of singularities'' is in
fact ``what one would expect from a quantum theory of gravity'' (see
p. 297 in \cite{Rovelli}).
An elementary support of such an intuitive
expectation
can be provided by the following schematic observable
 \be
 \Lambda(t)=
\left[ \begin {array}{cccc} 0&-1+i\,t&0&0\\
\noalign{\medskip}-1-i\,t&0&-1+i\,t&0\\
\noalign{\medskip}0&-1-i\,t&0&-1+i
\,t\\\noalign{\medskip}0&0&-1-i\,t&0
\end {array} \right]
\label{tyome}
 \ee
and by the inspection of its spectrum (see Figure \ref{uba2}).
As long as the matrix is Hermitian, its spectrum must be real.
Moreover, in the generic case (i.e., unless we impose a symmetry upon the matrix),
the spectrum {\em must\,} remain non-degenerate.
This is the reason why the levels avoid the crossing
(which would simulate the regularized Big Bounce).
In our example the proof of the phenomenon is elementary:
Up to a small vicinity of the ``Big Bang time'' $t^{(BB)}=0$
the matrix as well as its spectrum are dominated by their asymptotic
components which are strictly linear in $t$. One might even suspect that
the eigenvalues could cross due to an accidental symmetry emerging
at $t=0$ but such a symmetry is manifestly broken by
the $t-$independent component of the model.

We are going to show that
against all expectations,
the latter argument is not foolproof.
Admitting that it
need not {\em necessarily\,} lead to the wrong conclusions,
we will only show that the
Wang's and Stankiewicz's
alternative scenario \cite{nobounce}
may equally well be supported by an equally
elementary toy model.
The essence of such a claim is that the Hermiticity property
(cf. relation $\Lambda=\Lambda^\dagger$
satisfied by our toy-model matrix~(\ref{tyome}),
with the superscript $^\dagger$ marking the
matrix transposition plus complex conjugation)
depends on a mathematically motivated {\it a priori\,}
specification of the
inner product in our physical Hilbert space of states
\cite{Messiah}.


A deeper abstract foundation of our
``constructive scepticism''
concerning the genericity of the
Big Bounce
may be found in the
literature on
quantum mechanics using non-Hermitian operators
\cite{ali,Geyer,book,Carlbook}.
In this sense,
the common requirement of the
self-adjointness of the operators of observables $\Lambda(t)$
can be weakened and replaced by the condition of their
Hermitizability {\it alias\,} quasi-Hermiticity \cite{Geyer}.
In many non-Hermitian models, indeed, the Hermiticity may be restored
by the mere {\it ad hoc\,} amendment of the inner product
\cite{EPJP}.

In our present paper,
we will narrow the scope of the discussion to the
Big Bang and to the WDW equations.
Simultaneously, we will broaden the theoretical framework,
emphasizing that in the genuine Big Bang spatial-degeneracy context
it is necessary to replace the most common NSP mathematics
by its perceivably more complicated NIP amendment.
In a preparatory step
let us now return to the toy model (\ref{tyome})
and let us Wick-rotate the time $\, t \to - {\rm i\,}t$ and shift
the origin,
$\,t \to t-1$. The resulting new matrix
 \be
 Q(t)=
 \left[ \begin {array}{cccc} 0&-2+{t}&0&0\\\noalign{\medskip}-
{t}&0&-2+{t}&0\\\noalign{\medskip}0&-{t}&0&-2+{t}
\\\noalign{\medskip}0&0&-{t}&0\end {array} \right]
 \label{tyomed}
 \ee
is just a hiddenly Hermitian (i.e., via an amendment of the inner product,
Hermitizable) candidate for a toy-model observable representing,
in the context of quantum cosmology,
say, a potentially measurable
discrete
spatial grid \cite{grid,gridb,gridc,gridd,gride}.

\begin{figure}[h]                    
\begin{center}                         
\epsfig{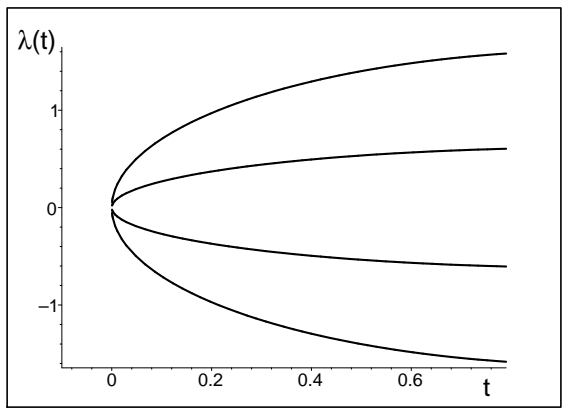}
\end{center}    
\vspace{2mm} \caption{The reality of spectrum
of non-Hermitian matrix (\ref{tyomed})
at not too large
$t\geq 0$.
 \label{uba3}
 }
\end{figure}

In essence, the latter example indicates that
the Big-Bang-type singularities {\em need not\,} necessarily be
smeared out by the quantization.
Indeed,
at the not too large values of the positive time parameter $t>0$
the spectrum of our manifestly non-Hermitian model (\ref{tyomed})
may be shown real and non-degenerate.
This is illustrated in Figure \ref{uba3}.
At  $t=0$ the spectrum
becomes degenerate and the matrix itself ceases to be
diagonalizable.

The latter simulation of the Big Bang singularity
is called exceptional point (EP) in mathematics \cite{Kato,symmetry}.
In the complementary context of physics, the spatial-grid interpretation
of the time-dependent eigenvalues
$\lambda_n(t)$ as sampled in Figure \ref{uba3}
enables us to speak about
the ``inflation period'' of the history of the related
hypothetical and
highly schematic
(i.e., four-point) quantized
Universe immediately after its birth.
Naturally, the corresponding internally consistent quantum theory must
be reformulated accordingly \cite{Geyer}.


\subsection{Stationary theory (non-Hermitian Schr\"{o}dinger picture, NSP)\label{wohn}}

In the pedagogically oriented and compact
review of the history of quantum mechanics \cite{nine}
the authors emphasized that there exists no
universal version
of quantum theory and that ``no formulation produces a
royal road to quantum mechanics''.
This explains an incessant emergence of the new versions of the theory
including its
recent ``non-selfadjoint-operator'' formulations \cite{book}.

Incidentally, the ``non-selfadjoint-operator'' characteristics
of these theories could be
misleading. As we already indicated, the mathematical concept of
non-selfadjointness
(or, in the shorthand terminology used by physicists, of non-Hermiticity)
is ambiguous. Covering, in various branches of physics,
{\em both\,} the generators
of the unitary evolution \cite{Geyer} {\em and\,} of the
non-unitary evolution \cite{Nimrod}.
It is necessary to emphasize that
only the former (i.e., unitarity-compatible)
meaning of the word `non-Hermiticity'
will be considered and taken into account in what follows.

The disambiguation really deserves the early notice because
the difference is often less clear in applications.
Also the
formulation of the {\em physical\,} background of the
problems
happens to suffer of ambiguities. The details
will be discussed later (see, first of all,
the introduction in the problem as given in Appendix).
Now, let us
only
repeat that the questions
which we are going to discuss
will have their origin in the field of
quantum gravity \cite{Thiemann}. In this broad context
our attention will be paid, first of all, to the possible role played by
the WDW equation and to the questions of
physics near the Big Bang (BB) singularity.

\subsection{\label{ztseci2}Stationary Wheeler-DeWitt equation}

In the stationary case the WDW problem becomes
formally equivalent to the Klein-Gordon (KG)
problem known in the relativistic quantum mechanics \cite{aliWDW}.
In their
simplest versions, both
of these problems may be characterized,
in suitable units, by the linear differential equation.
Thus, in the KG case  (where the suitable units are $\hbar=c=1$
and where one omits, for the sake of simplicity,
the electromagnetic field)
we have, for example,
 \be
 \left (
 \frac{\partial^2}{\partial t^2}+D^{(KG)}\right
 )\,\psi^{(KG)}(\vec{x},t)=0\,,
 \ \ \ \ \ \ D^{(KG)}=-\triangle +m^2\,.
  \label{kgha}
 \ee
The
kinetic energy is represented here by the elementary Laplacean
$\triangle$, and the dynamics can be
maximally reduced to the mere scalar mass term
which may be made position dependent,
$m^2=m^2(\vec{x})$.

In the simplest non-stationary WDW
model the analogue of the mass term would be a time-dependent function
(cf., e.g., section Nr. 3.5 of review \cite{ali} for further references).
The KG-WDW analogy enables us to use the same mathematical tools.
The relevant literature is fairly extensive but
for our present purposes it is sufficient to cite
the paper by Feshbach and Villars (\cite{FV}, cf. also Ref.~\cite{WDWja})
in which the change of
variables
 \be
 \psi^{(WDW)}(\vec{x},t) \ \ \to \ \
 \br \vec{x}|\psi^{(FV)}(t)\kt=
 \left (
 \ba
 {\rm i}\partial_t \psi^{(WDW)}(\vec{x},t)\\
 \psi^{(WDW)}(\vec{x},t)
 \ea
 \right )\,
  \label{assi}
 \ee
was shown to lead to a replacement of the hyperbolic partial
differential Eq.~(\ref{kgha}) by the
Schr\"{o}dinger-like parabolic equation
for the two-component wave function (\ref{assi}),
 \be
 {\rm i} \frac{\partial}{\partial t} \,
 \,|\psi^{(FV)}(t)\kt=G_{(FV)}(t)\,|\psi^{(FV)}(t)\kt
  \,.
 \ee
This equation can be interpreted as
controlling the unitary evolution
of the system via the generator {\it alias\,} FV Hamiltonian
 \be
 \ \ \
 G_{(FV)}(t)=
 \left (
 \begin{array}{cc}
 0&D(t)\\
 I&0
 \ea
 \right )\neq G_{(FV)}(t)\,.
  \label{SEFekg}
 \ee
Such an operator is, in the FV Hilbert space $${\cal H}^{(FV)}= {\cal
L}^2(\mathbb{R}^3)\,\bigoplus\,{\cal   L}^2(\mathbb{R}^3)$$
manifestly
non-Hermitian,
$ G_{(FV)} \neq G_{(FV)}^\dagger$.
Pauli with
Weisskopf \cite{PW} noticed that
the same operator can in fact be treated as
self-adjoint
with respect to another, indefinite inner product,
 \be
 \br \psi_1|\psi_2\kt \ \ \to \ \
 \left ( \psi_1,\psi_2 \right )_{(Krein)}
 =\br \psi_1|{\cal P}_{(FV)}|\psi_2 \kt
\,.
 \label{tran}
 \ee
i.e., that it is self-adjoint in another, {\it ad hoc\,} Krein
space. In the modern terminology
one would say that this operator is non-Hermitian but
${\cal PT}-$symmetric \cite{Carl}.

A decisive progress achieved under the stationarity assumption
$ G_{(FV)} \neq G_{(FV)}(t)$
(or, more precisely, after
its generalized form called
quasi-stationarity assumption)
is due to
Mostafazadeh. In his papers \cite{aliWDW,[150]}
he imagined that the FV pseudometric
${\cal P}$ could be replaced by
the positive definite metric $\Theta_{(stationary)}$
converting the Krein-space physics
(in which, during evolution, the
usual norm is not conserved)
into the
fully standard and norm-conserving Hilbert-space physics.
In essence, just a straightforward change of the
inner product was needed,
 \be
 \left ( \psi_1,\psi_2 \right )_{(Krein)}\ \ \to \ \
 \left ( \psi_1,\psi_2 \right )_{(Mostafazadeh)}
 =\br \psi_1|\Theta_{(stationary)}\,|\psi_2 \kt\,.
 \label{trans}
 \ee
This opened the way towards a consistent
picture of unitary physics in which the stationary Hamiltonian
$ G_{(FV)} =H_{(FV)}$
controls the NSP quantum evolution which is,
with respect to the amended inner product
(\ref{trans}), unitary.

After either the KG or the WDW interpretation of Eq.~(\ref{kgha})
in stationary case,
the Hilbert-space metrics in (\ref{trans}) can be given a
formal block-diagonal-operator structure
 \be
 \Theta_{(stationary)}=
 \left (
 \begin{array}{cc}
 1/\sqrt{D}&0\\
 0&\sqrt{D}
 \ea
 \right )\,.
 \label{thassi}
 \ee
This leads to the first quantization of both of these systems.


\section{Fine-tuned nature of the quantum Big Bang
\label{dodada}}

The conventional mental operation called ``quantization of the classical theory''
does really very naturally lead to the conclusion that the singularity
gets ``smeared our'' near $t \approx 0$
due to ``quantum effects'' \cite{AshtekarBi,AshtekarBib} (see also
the four-by-four Hermitian matrix (\ref{tyome})).
In our text we pointed out that the support
of such a regularization hypothesis
is only unavoidable in the conventional ``textbook'' quantum mechanics.
In a more general, hiddenly Hermitian theory such an assumption is
artificial and unfounded (cf. Appendix or toy model (\ref{tyomed})).
Once one overcomes the mental barrier
one reveals that the
inner product
may
start playing the central descriptive role.


\subsection{The $N-$grid-point toy model of kinematics}

In the literature the manifestly non-Hermitian
but Hermitizable
Wheeler-DeWitt equation has only been considered
in the stationary (or, better, quasi-stationary)
mathematical NSP regime (cf. \cite{ali} or section \ref{wohn} above).
In section \ref{seci2} we will turn attention to
the conceptual necessity of keeping the WDW-related
Hilbert space time-dependent.
In an overall context of
the canonical quantization of gravity
we have to be prepared to address, therefore,
a number of purely technical questions and tasks.

In the first one the point-like
Big Bang must be made compatible with a
consequent theoretical unitary-evolution scenario.
Thus, we have to
complement the abstract argumentation of section \ref{hown}
by a
detailed description of a suitable concrete
toy model. In the model
the measurable
values of the spatial grid points (say, the necessarily real
and time-dependent values $q_j(t)$ with $j=1,2,\ldots N$)
will have to be assumed obtainable, in principle at least, as eigenvalues
of a suitable
hiddenly Hermitian geometry-representing
``effective kinematical input'' operator (say, $Q^{(N)}(t)$).

Secondly, we have to keep in mind that in a way indicated by
our four-by-four matrix (\ref{tyomed})
we will assume that the general $N$ by $N$ matrix $Q^{(N)}(t)$
will still be real and tridiagonal.
Indeed, in a way explained in~\cite{recurrently}
the reality and tridiagonality is an important merit of {\em any\,}
candidate for an observable because it enables
one to
construct
the metric algebraically, in recurrent manner.
In this sense we may recall the existing results in linear algebra \cite{tridiagonal}
and choose the one-parametric family of our $N$ by $N$ toy-model
``effective kinematics'' as follows,
 \be
 Q^{(N)}(z)=
  \left[ \begin {array}{cccccc}
   -i\,(N-1)z&- \sqrt{N-1}&0&0&\ldots&0
 \\\noalign{\medskip} - \sqrt{N-1}&-i\,(N-3)z&-\sqrt{2(N-2)}&0&\ddots&\vdots
 \\\noalign{\medskip}0&-\sqrt{2(N-2)}&-i\,(N-5)z&\ddots&\ddots&0
 \\\noalign{\medskip}0&0&-\sqrt{3(N-3)}&\ddots&-\sqrt{2(N-2)}&0
 \\\noalign{\medskip}\vdots&\ddots&\ddots&\ddots&i\,(N-3)z&-\sqrt{N-1}
 \\\noalign{\medskip}0&\ldots&0&0&-\sqrt{N-1}&i\,(N-1)z\end {array}
 \right]\,.
 \label{ponasem}
 \ee
The non-triviality of this matrix and the arbitrariness of its dimension $N$
in combination with
its non-numerical
tractability \cite{minimal} will enable us to
show how the requirement of the existence of the quantum Big Bang
singularity becomes supported by a consistent
reconstruction of the related
physical time-dependent Hilbert-space metric.
As long as $z=z(t)$ can be any suitable function of time
we may restrict our considerations to the interval of
$z \in (-1,1)$ in the interior of which the
grid-point-coordinate spectrum of $ Q^{(N)}(z)$
remains non-degenerate, real and discrete, and at the boundaries of which
one can visualize the realization of the Big Bang.
Thus, after the simplest choice of $z(t)=-1+t$ we obtain
an immediate $N-$level analogue of the
graphical evolution pattern
of Eq.~(\ref{uba3})
where we have
$N=4$.

One of the main constraints imposed upon our
toy-model ``geometry operator''~(\ref{ponasem}) is its
compatibility with the unitarity of the
quantum evolution, i.e., with the existence of
the Hilbert-space
metric. Naturally, the process of the evolution of the
corresponding schematic Universe will have to
start at the Big Bang single-point-degeneracy singularity
which is such that
 \be
 \lim_{t \to 0^+} q_n(z(t)) = q_1(z(0))\,,
 \ \ \ \ \ n = 1,2, \ldots, N\,
 \label{degere}
 \ee
On the technical level,
one can really speak about a challenge because
even the purely formal
construction of a highly schematic
``Big-Banging'' model of the quantum Universe
must remain compatible with the basic theoretical requirement of
compatibility between the
kinematical
input information (\ref{degere}) and
the dynamical
input information
as represented by the WDW Hamiltonian operator.
The details will be discussed below. For the time being let us only
assume that
with the kinematical spatial-grid input (\ref{degere})
adapted
to {\em any\,}
phenomenological requirements,
the dynamics
of the WDW-related Universe
will remain reflected by
a suitable non-stationary form of the
operator $D$ in its form entering the non-stationary
analogue of the stationary (or, if you wish, adiabatic)
form (\ref{thassi}) of the
WDW Hilbert-space metric.

\subsection{The fine-tuned nature of the Hilbert-space metric $\Theta(z)$
\label{ambo}} 

One of our most important WDW-related
model-building tasks can be seen in the
generalization of the qualitative
and consistent picture of the quantum
Big Bang singularity as mediated by
its $N=4$ grid-point realization via
Eq.~(\ref{tyomed}) above (cf. also Figure \ref{uba3}).
In such project we encounter
the two main technical obstacles. The first one
lies in the necessity of the guarantee of
the existence of the metric $\Theta$ at all
times $t>0$
(i.e., in our model, at all of the sufficiently small
positive times) {\em up to the very Big-Bang
birth-of-the-Universe EP limit\,}
$t \to 0^+$.
In our toy model, due to its
exact solvability \cite{tridiagonal},
such a guarantee will have an
exact, non-numerical form.

The way of circumventing the second technical obstacle
(viz., the necessity of a guarantee that the Hilbert-space metric remains,
at all of the relevant times, non-singular and positive definite)
is equally difficult to find. In our model we shall see that
for the model in question this goal can be achieved
by the non-numerical means as well.

The respective solutions of both of the above-mentioned problems
are closely interrelated.
Their essence can
be identified with the necessity of
the coexistence
of the singularity in the grid
with the
singularity-free
nature of the metric $\Theta(z)$.
The most universal
approach to this problem has been promoted by Scholtz et al
\cite{Geyer} who proposed to use a {\em complete\,} information about
the set of the observables $\Lambda_1(z)$,
$\Lambda_2(z)$, \ldots.
Such an ``extreme'' model-building strategy yielded
a unique physical metric $\Theta^{(N)}(z)$.
In principle, its applicability is strongly $N-$dependent of course.
Thus, our methodical considerations will only concern
the systems with the smallest dimensions.

\subsubsection{The eligible Hilbert-space metrics at $N=2$}

At $N=2$, the grid-point operator (\ref{ponasem}) reads
 \be
 Q^{(2)}(z)=\left[ \begin {array}{cc} -iz&-1
 \\\noalign{\medskip}-1&iz\end {array}
 \right]\,, \ \ \ \ \ \ z \in (0,1)\,.
 \label{ha2}
 \ee
With the four real parameters $a,c,d$ and $\chi \in (0,2\pi)$,
with, for the sake of definiteness, positive $z \in (0,1)$, and
with the general ansatz
 \be
 \Theta^{(ansatz)}(a,c,d,\chi)=\left[ \begin {array}{cc} a&c\,e^{-{\rm i}\chi}
 \\ \noalign{\medskip}c\,e^{{\rm i}\chi}&d\end {array}
 \right]
 \label{anss}
 \ee
for the Hilbert-space metric
the condition of quasi-Hermiticitiy degenerates to the two elementary
relations,
 $$
 d=a=z^{-1}\,
 c\,\sin \chi\,.
 $$
Without any loss of generality we may set $c=z$ and evaluate the
eigenvalues of matrix of Eq.~(\ref{anss}),
 $$
 \lambda_\pm = \sin \chi \pm z\,.
 $$
Thus, this matrix may be declared acceptable as a metric if and only if it is
positive definite, i.e., if and only if
 \be
 \sin \chi > z\,.
 \label{compa}
 \ee
This relation clearly indicates that
near the EP limit $z \to 1$ the range of variability of
the admissible parameter $\chi$
(numbering the admissible Hilbert-space metrics)
becomes extremely narrow.
Moreover,
whenever
the dynamics-controlling parameter
$z$ moves closer to the EP singularity
the interval quickly shrinks
so that our choice of the metric must be,
in the Big Bang vicinity, very precisely
``fine-tuned''.

Formula (\ref{compa}) gets further simplified when we reparametrize the
strength of the non-Hermiticity $z=\sin \beta\ $ in terms of the new
variable $\beta \in (0,\pi/2)\ $. Now, the Hermitian limit
corresponds to $\beta=0$ while the singular EP
(or, if you wish, Big Bang or Big Crunch)
extreme is reached
at $\beta=\pi/2$. Ultimately, formula
 \be
 \Theta(\beta,\chi)=\left[ \begin {array}{cc}
 \sin \chi &e^{-{\rm i}\chi}\,\sin \beta
 \\ \noalign{\medskip}e^{{\rm i}\chi}\,\sin \beta&\sin \chi \end {array}
 \right]\,,
 \ \ \ \ \ \ \
 \chi \in (\beta,\pi-\beta)\,
 \label{30}
 \ee
defines, up to an inessential overall factor, all of the eligible
correct metric operators at $N=2$.

\subsubsection{$N=3\,$ and the requirement of positivity\label{Bchpetka}}


Once we move to the next geometry operator (\ref{ponasem}) with $N=3$, the
general ansatz for the metric may be reduced to a six-parametric Hermitian
matrix
 \be
 \Theta=\left[ \begin {array}{ccc} a&b{e^{i\phi}}&c{e^{i\chi}}
 \\\noalign{\medskip}b{e^{-i\phi}}&f&b{e^{i\phi}}\\\noalign{\medskip}c{
 e^{-i\chi}}&b{e^{-i\phi}}&a\end {array} \right]\,.
 \label{orino}
 \ee
This reveals that the construction of the metric remains a
purely routine linear-algebraic problem.
At the same time,
the
weakness of the construction
is found to lie in the less easy
determination of the domain of parameters for
which the metric operator $\Theta$ remains positive definite.
Although
the domain of
positivity of the metric
is still implicitly defined by the $N=3$ secular determinant and by the
relation
 $$
 {{\it \lambda}}^{3}+ \left( -f-2\,a \right) {{\it \lambda}}^{2}+ \left(
 -2\,{b}^ {2}-{c}^{2}+{a}^{2}+2\,fa \right) {\it \lambda}
  +{c}^{2}f-f{a}^{2}+2\,a{b}^{2}-2\,{b}^{2}c\cos
 \left( 2\, \phi-\chi \right)=0\,
 $$
the $N=3$ analogue of the $N=2$ formula (\ref{30})
would be complicated for an explicit display.
The task still remains
non-numerical because the secular polynomial remains
linear in the parameters $b^2$, $f$ and/or $\cos \left( 2\, \phi-\chi
\right)$. This still leaves the determination of
the range of the admissible parameters straightforward.
A typical sample of such a determination is provided by Figure \ref{hecuba}.

\begin{figure}[h]                    
\begin{center}                         
\epsfig{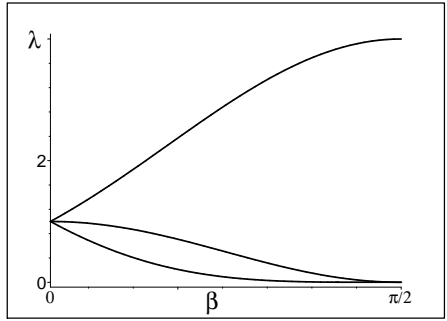}
\end{center}    
\vspace{2mm} \caption{Eigenvalues of our singularity-free metric (\ref{orino})
as functions of one of the parameters at
$N=3$.
 \label{hecuba}
 }
\end{figure}

In this illustrative picture we see that
the
eigenvalues of the metric remain real and non-degenerate
in a large
interval of one of the dynamical $N=3$ parameters $\beta$.
In a small
vicinity of the singular EP/BB limit $\beta\to \pi/2$
we may deduce that the rank of the metric becomes
approximately equal to one.
The picture even shows the confluence of the eigenvalues
of the metric
in the trivial-metric Hermitian-system limit
$\Theta \to I$, i.e., very far from the EP/BB dynamical regime.

Naturally, the technical difficulties will grow with the dimension.
At the larger $N$
the construction has to be given an
alternative, purely graphical form.
This strategy has been used in paper \cite{annalsix3}
where it has been
shown that the use of the graphical method remains feasible
even for the higher-order secular polynomials
(cf. Figures Nr. 16 and 17 in {\it loc. cit.}).
Stil, one has to expect that at the truly large matrix dimensions
the construction becomes purely
numerical.

\subsection{Candidates for the other observables}

For any given non-Hermitian grid-point operator $Q^{(N)}(z)$ with the
real and nondegenerate spectrum $\{q_n(z)\}$ one can construct the
arbitrarily normalized eigenvectors,
 \be
 Q^{(N)}(z)
 \,|n(z)\kt = q_n(z)\,|n(z)\kt\,,
 \ \ \ \
 n = 1,2, \ldots, N\,.
 \ee
For the same spectrum, the
arbitrarily normalized double-bra-marked left eigenvectors
may be also defined
as the standard right eigenvectors
of a Hermitian conjugate operator,
 \be
 \left [Q^{(N)}(z)
 \right ]^\dagger\,|n(z)\kkt = q_n(z)\,|n(z)\kkt\,,
 \ \ \ \
 n = 1,2, \ldots, N\,.
 \label{dirig}
 \ee
It is easy to deduce
that $\bbr m|n\kt = 0$ for $m \neq n$. In the generic case the
overlaps $\bbr m|m\kt$ will be real and non-vanishing.
Whenever $N$ is finite, the resulting
biorthogonal basis
can be used in a
generalized spectral representation of the operator
 \be
 Q^{(N)}(z)=\sum_{n=0}^{N-1}\, |n(z)\kt \,\frac{q_n(z)}{\bbr n(z)|n(z)\kt}\,\bbr n(z)|\,.
 \ee
We may conclude that
the general
(though not necessarily invertible or positive definite)
$N-$parametric Hilbert-space metric can be then defined by formula
 \be
 \Theta^{(N)}=\sum_{n=0}^{N-1}\, |n\kkt \,\kappa_n\,\bbr n|\,.
 \label{wright}
 \ee
The parameters $\kappa_n$ must be all real. The acceptability of
the matrix in the role of the physical Hilbert-space metric
(i.e., the necessary
invertibility and positivity properties) are
then guaranteed if and only if $0<\kappa_n<\infty$ at all $n$
\cite{SIGMAdva}.
In such a setting, one can easily
use an analogous generalized spectral representation to
define also any other operator of an acceptable
quantum observable.

\section{\label{seci2}
Mathematics: Non-Hermitian interaction picture (NIP)}

Naturally,
the (quasi-) stationarity
restriction becomes, in the WDW case, hardly acceptable,
especially if one tries to deal with the quantum dynamics near
a singularity like Big Bang.
In such a case, a much deeper modification of the formalism
of the non-Hermitian quantum theory is needed.

After one decides to relax the assumptions of stationarity,
an increase of the complexity of the system of equations
is partially compensated by the clarification of several
conceptual problems. In this sense our main
methodical recommendation is that
in analogy with the Hermitian interaction picture of textbooks,
one still keeps in mind the
necessity of description of the dynamics in terms of both the
operators and wave functions. In other words, it is necessary to
avoid several existing and wide-spread misunderstandings
which can be found in the current literature.
Paradoxically, the root
of these misunderstandings may be seen in an insufficiently careful
use of the terminology (see, e.g.,
the explanatory ``Rosetta-stone-like'' Table Nr. 1 in \cite{NIP}).
Indeed, once we replace a stationary
NSP model by its non-stationary IP and/or NIP alternative and extension,
the concept of quantum Hamiltonian
ceases to be unique and adequate.

\subsection{\label{ztseci3}Non-stationary quantum systems}

In the non-stationary quantum theory the use of the
time-dependent metric is known to lead to the loss of the
unitarity of the evolution or to the loss of the observability of the
NSP Hamiltonian \cite{ali}. In fact \cite{which}, the
puzzle is
artificial and purely terminological.
The problem disappears when one employs the non-Hermitian
version of the Dirac's interaction picture (NIP, \cite{NIP}).

\subsubsection{Evolution law for the NIP ket
vectors\label{neramo}}

In
the non-stationary non-Hermitian cases
there is no need of the observability of the generator
of the evolution of the ket-vectors~\cite{Heisenberg,cinani,FringMou,Luiz}.
Easily, the stationary
version of the Dyson map (\ref{lumapo}) can be replaced by its
time-dependent generalization
 \be
 |\psi^{}(t)\pkt=\Omega(t)\,|\psi^{}(t)\kt
 \in {\cal H}^{(T)}\,,
 \ \ \ \ \ \ |\psi^{}(t)\kt \in {\cal H}^{(F)}\,.
 \label{mapoge}
 \ee
In ${\cal H}^{(F)}$, similarly, Schr\"{o}dinger
Eq.~(\ref{SETdys}) acquires the form
 \be
 {\rm i} \frac{\partial}{\partial t}
 \,|\psi^{}(t)\kt=G_{}(t)\,|\psi^{}(t)\kt\,
  \label{SEFip}
 \ee
in which the generator is
only one of the two unobservable components of the
observable
instantaneous-energy operator
 \be
  H(t) =G(t) + \Sigma(t)
 \,.
  \label{2.3}
 \ee
Only the sum will be called
Hamiltonian in what follows.
The other component of the Hamiltonian
can be defined
directly in terms of the Dyson map,
 \be
 \Sigma(t)={\rm i} \Omega^{-1}(t)\,\dot{\Omega}(t)\,,
 \ \ \ \ \
 \dot{\Omega}(t)=
 \frac{d}{dt} \,
 \Omega(t)\,
 \label{defsig}
 \ee
(see \cite{timedep,SIGMA,NIP}
for details).

In the unitary-evolution case the observable
version of the non-Hermitian
but Hermitizable
Hamiltonian (\ref{2.3})
is connected with its self-adjoint partner by formula,
 \be
 H_{}(t)= \Omega_{}^{(-1)}(t)\,
 \mathfrak{h}_{(NSP)}(t)\,\Omega_{}(t)\,.
 \label{udagobs}
 \ee
In ${\cal H}^{(F)}$ operator (\ref{udagobs}) has the property of
quasi-Hermiticity,
 \be
  H^\dagger(t)\, \Theta(t)=
  \Theta^{}(t)\,H_{}(t)
 \,,
 \ \ \ \
 \Theta(t)=\Omega^\dagger(t)\,\Omega(t)\,.
 \label{dagobs}
 \ee
In an internally consistent theory
of a unitary (or hiddenly unitary) quantum system
the Hamiltonian still has to have
the real and discrete
spectrum representing the instantaneous
(but still observable)
bound-state energies.

It is unfortunate that in the literature, only too may people
assign the name of a Hamiltonian also
to {\em both\,} of the other operators
$G(t)$
and $\Sigma(t)$, neither of which represents
an observable quantity \cite{which,SIGMA,FringMou}.
We prefer calling operator
$G(t)$ a ``generator''
(which does not represent an observable while still
controlling and generating the evolution of the IP/NIP wave functions).
In parallel,  we would also propose calling operator
$\Sigma(t)$, say, a ``Coriolis force''.

\subsubsection{Evolution law for the NIP bra vectors\label{paramo}}

Equation~(\ref{mapoge}) has a dual-space alternative
 \be
 |\psi^{}(t)\pkt
  =\left [\Omega^\dagger(t)\right ]^{-1}|\psi_\Theta(t)\kt
 \in {\cal H}^{(T)}\,,
 \ \ \ \ \ \ |\psi_\Theta(t)\kt \ \equiv \ \Theta(t)\,|\psi(t)\kt
  \in {\cal H}^{(F)}\,.
 \label{dumapo}
 \ee
This enables us to treat the new states
$|\psi_\Theta(t)\kt \ \equiv \ \Theta(t)\,|\psi(t)\kt$
as solutions of another Schr\"{o}dinger equation in ${\cal
H}^{(F)}$ \cite{timedep,SIGMA},
 \be
 {\rm i} \frac{\partial}{\partial t} \,|\psi_\Theta(t)\kt
 =G^\dagger(t)\,|\psi_\Theta(t)\kt\,.
  \label{d2.3}
 \ee
The process of solution of the two
Schr\"{o}dinger equations is maximally economical.
The key merit of this recipe
(see also more commentaries in \cite{NIP}) is that it circumvents
the necessity of
the technically much
more complicated direct construction of the metric
as used, e.g., in papers
\cite{FringMou,Wang,Bila,Bilab,FrFrith}.

The present version of the process
must be
initiated by the specification
of the respective states
$|\psi(t)\kt$ and $|\psi_\Theta(t)\kt$ at
$t=t_i=0$.
Thus, equations~(\ref{SEFip}) and
(\ref{d2.3}) have to be
complemented by the specification of the
initial values represented by the kets $|\psi(t_i)\kt$ and
$|\psi_\Theta(t_i)\kt$.
Naturally, such
values must obey constraints~(\ref{mapoge}) and (\ref{dumapo}) at
$t=t_i=0$.
This, in turn, is
closely connected with the
experiment and with the preparation of the system
in question.

\subsection{\label{xseci2}
Non-Hermitian operators in interaction picture}

It is well known that even
in the conventional Hermitian version of IP,
the Coriolis-force operators obey the Heisenberg-type equations.
These equations
control the evolution of {\em every\,} relevant
operator of an observable.

In the non-Hermitian NIP formalism the role of $\Sigma(t)$
is analogous.
In both of the IP and NIP cases, the ultimate goal of the
theory lies in the derivation of the
predictions of the results of measurements.
In our present version of the recipe, this merely requires
the evaluation of the overlaps
 \be
 \br \psi^{}_\Theta(t_f)\,|Q_{}(t_f)\, |\psi^{}(t_f) \kt\,.
 \label{redysMEAS}
  \ee
Due to the
identity
 \be
 {\rm i}\,\frac{\partial}{\partial t} \,
 \Theta(t)= \Theta(t)\,\Sigma(t) - \Sigma^\dagger(T)\, \Theta(t)\,
 \label{identity}
 \ee
or due to its alternative version (cf. Eq.~(\ref{dagobs})),
 \be
 {\rm i}\,\frac{\partial}{\partial t} \,
 \Theta(t)= G^\dagger(t)\, \Theta(t)-\Theta(t)\,G(t) \,
 \label{2.6}
 \ee
the NIP formalism is internally consistent, indeed.
At the same time, one has to keep in mind that the
operators of the IP or NIP observables are manifestly time-dependent
and that their time-dependence is not arbitrary.

\subsubsection{Evolution law for the density matrices
\label{huramo}}

In the non-Hermitian
but unitary
pure-state quantum systems
of our present interest
the state
is defined by {\em a pair\,} of the ket vectors,
i.e., by the
projectors
 \be
 \pi_{\psi,\Theta}(t)=|\psi(t)\kt \,\frac{1}{\br
 \psi_\Theta(t)\,|\psi(t)\kt}\,\br \psi_\Theta(t)|\,.
 \ee
Alternatively, one can speak about the
non-Hermitian density matrix
 \be
 \widehat{\varrho}(t)=\sum_{k}|\psi^{(k)}(t)\kt
 \,\frac{p_k}{\br \psi^{(k)}_\Theta(t)\,|\psi^{(k)}(t)\kt}\,\br
 \psi^{(k)}_\Theta(t)|
 \,,
 \ \ \ \ \ \
 \sum_{k} p_k=1\,.
 \ee
Due to
Eqs.~(\ref{SEFip}) and (\ref{d2.3}), this operator has to obey
the specific evolution equation
 \be
 {\rm
 i\,}\partial_t\, \widehat{\varrho}(t)= G(t)\,\widehat{\varrho}(t)
 -\widehat{\varrho}(t)\,G(t)\,
 \ee
which opens the way towards the formulation of
quantum statistics in non-Hermitian Liouvillean
picture \cite{NIP}.

\subsubsection{The evolution of
observables\label{koramo}}

The
requirement
 \be
 Q_{}^\dagger(t)\,\Theta(t)=\Theta(t)\,Q(t)
 \label{nenene}
 \ee
guarantees the observability status of any
operator $Q_{}(t)$. This relation is equivalent,
due to Eq.~(\ref{dadag}), to
the NSP
Hermiticity of $\mathfrak{q}_{}(t) $ in ${\cal H}^{(T)}$
since
 \be
 Q_{}(t)= \Omega_{}^{(-1)}(t)\,
 \mathfrak{q}_{(NSP)}(t)\,\Omega_{}(t)\,
 \label{redadag}
 \ee
The Heisenberg-type evolution
equation follows,
 \be
 {\rm i\,}\frac{\partial}{\partial t} \,{Q}_{}(t)=
  Q(t)\,\Sigma(t) -\Sigma_{}(t)\,Q(t)
 +K(t)\,,
 \ \ \ \ \ K(t)=\Omega_{}^{(-1)}(t)\,
 {\rm i\,}\dot{\mathfrak{q}}_{(NSP)}(t)\,\Omega_{}(t)\,.
 \label{beda}
 \ee
It is recommendable to assume that the partial derivatives
$\dot{\mathfrak{q}}_{(NSP)}(t)$ vanish so that the related operator $K(t)$ would
be vanishing as well, making the process of the solution of
Eq.~(\ref{beda}) user-friendlier.

Given the generator $G(t)$, the choice of the Coriolis force
$\Sigma(t)$ is far from arbitrary. First of all, it is constrained by
the experiment-related
initial state vectors.
Secondly, it must be compatible with its relation (\ref{2.3})
to the initial
instant energy $H^{(NIP)}(t_i)$
and to its evolution law
 \be
 {\rm i\,}\frac{\partial}{\partial t} \,{H}^{(NIP)}(t)=
  H^{(NIP)}(t)\,\Sigma^{(NIP)}(t)
  -\Sigma^{(NIP)}(t)\,H^{(NIP)}(t)+K^{(NIP)}(t)
 \label{bedakat}
 \ee
or, equivalently,
 \be
 {\rm i\,}\frac{\partial}{\partial t} \,{H}^{(NIP)}(t)=
     G^{(NIP)}(t)\, H^{(NIP)}(t)
     -H^{(NIP)}(t)\,G^{(NIP)}(t)+K^{(NIP)}(t)\,.
 \label{rebedakat}
 \ee
Next, one will also frequently decide to accept the important simplification
obtained for the
vanishing NSP-time-derivative operators
 $$
K^{(NIP)}(t)=\Omega_{}^{(-1)}(t)\, {\rm
i\,}\dot{\mathfrak{h}}_{(NSP)}(t)\,\Omega_{}(t)\,.$$
As long as
$\Sigma^{(NIP)}(t) = H^{(NIP)}(t) - G^{(NIP)}(t)$ there remains no
freedom left. In particular, as long as we
have the definition,
 \be
 {\rm i} \frac{\partial}{\partial t} \,\Omega^{(NIP)}(t)\kt=
\Omega^{(NIP)}(t)\,\Sigma^{(NIP)}(t)\,,
  \label{reniSEFip}
 \ee
the only ambiguity of $\Omega^{(NIP)}(t)$ is contained in its
initial-value specification.

\section{The construction of non-stationary WDW Universe
admitting Big Bang\label{seci3}}


In our present study of applicability
of the NIP approach to the various
models in cosmology
we felt particularly interested
in a guarantee of the Big Bang degeneracy property
 \be
 \lim_{t\to 0^-}q_j(t) = 0\,,\ \ \ \ j = 0, 1, \ldots\,
  \label{leaq}
 \ee
which, in the formal context of quantum mechanics, prescribes
and restricts the
behavior of certain time-dependent eigenvalues $q_j(t)$
of a suitable operator
characterizing the spatial geometry
(or at least the size)
of the Universe. Sampled,
say, by $Q(t)$ of Eq.~(\ref{redadag}),
or by $\widetilde{\Lambda}(t)$ of Eq.~(\ref{tyomed}),
with the spectrum as sampled in Figure \ref{uba3}.
For this purpose
let us now return to some less general, simplified WDW models.

\subsection{The evolution of the WDW ket-vectors}

Even in the non-stationary cases, many
KG and WDW models remain formally equivalent.
For this reason let us now return to Eq.~(\ref{kgha})
replaced by its non-stationary generalization
 \be
 \left (
 \frac{\partial^2}{\partial t^2}+D(t)\right
 )\,\psi^{(WDW)}(\vec{x},t)=0\,,
 \ \ \ \ \ \ D(t)=-\triangle +m^2(\vec{x},t)\,.
  \label{kghage}
 \ee
Using the same amendment of the wave functions as before,
 \be
  \br \vec{x}|\psi^{(NIP)}(t)\kt=
 \left (
 \ba
 {\rm i}\partial_t \psi^{(WDW)}(\vec{x},t)\\
 \psi^{(WDW)}(\vec{x},t)
 \ea
 \right )\,
  \label{prassi}
 \ee
we are able to replace Eq.~(\ref{kghage}) by
an analogue of Eq.~(\ref{SEFip}),
i.e., by the correct NIP
Schr\"{o}dinger equation
 \be
 {\rm i} \frac{\partial}{\partial t} \,|\psi^{(NIP)}(t)\kt=
\left (
 \begin{array}{cc}
 0&D(t)\\
 I&0
 \ea
 \right )\,|\psi^{(NIP)}(t)\kt\,.
  \label{niSEFip}
 \ee
Here, the spectrum of the WDW generator $ G_{(NIP)}(t)$ need not be real
of course (see, for example, an elementary illustrative example
as given in \cite{SIGMA}).

\subsection{The evolution of the WDW bra-vectors}

It is obvious
that the time-dependence of the
metric $\Theta(t)$ may be highly sensitive to its
initial value at $t=t_i$ \cite{FringMou,Wang,Bila,IJTP}.
Unfortunately, the direct analysis of this dependence via
the solution of Eq.~(\ref{2.6}) is complicated.
For this reason we recommended, in \cite{NIP},
to follow the guidance by papers \cite{Heisenberg,IJTP}
and to circumvent the solution of the auxiliary {\em operator\,}
evolution Eq.~(\ref{2.6})
(which was characterized, in
\cite{FringMou}, as the ``time-dependent
quasi-Hermiticity relation'')
and to solve the second Schr\"{o}dinger equation
(for the mere bra-{\em vectors}) instead.

This leads to the implementation of the NIP recipe with
the evolution of
 \be
 |\psi_\Theta^{(NIP)}\kt = \Theta(t)\,|\psi^{(NIP)}\kt
 \ee
controlled by Schr\"{o}dinger
Eq.~(\ref{d2.3}),
 \be
 {\rm i} \frac{\partial}{\partial t} \,|\psi_\Theta^{(NIP)}(t)\kt=
\left (
 \begin{array}{cc}
 0&I\\
 D^*(t)&0
 \ea
 \right )\,|\psi_\Theta^{(NIP)}(t)\kt\,.
  \label{niSEFiq}
 \ee
Here, it is necessary to emphasize that once we identified
the NIP generator $G(t)$
with
the WDW generator in Eqs.~(\ref{niSEFip}) and (\ref{niSEFiq}), we made,
in effect, a certain highly nontrivial decision.
It has two aspects. In the phenomenological context such a
decision implies
that the WDW generator {\em does not\,} represent an observable.
We believe that there are all reasons for such a preference,
especially in the context of the possible quantization of gravity
because in such a context the WDW eigenstates are usually treated as
a means of specification of the Hilbert space rather than as the
observable states which would be
directly
connected with the energy \cite{Rovelli}.

\subsection{Reconstruction of the metric $\Theta(t)$ from the generator $G(t)$
\label{seci3b}}

In the NIP framework it is sufficient
to admit that only the sum (\ref{2.3})
of
the generator $G(t)$ and of the Coriolis-force
$\Sigma(t)$ (of a purely kinematical origin)
can be interpreted as the observable Hamiltonian.
In such a non-stationary NIP scenario, several open questions
emerge and
have to be resolved of course.

\subsubsection{Big Bang rendered possible by the time-dependence of the metric}

Let us now accept the
model-building
strategy in which one is given
the kinematical input
operator $G(t)$.
Then, the general non-Hermitian interaction picture can be declared
exceptional because only this picture
is in fact a candidate for a realization of
the quantum Big-Bang-like phase transitions
via a unitary evolution process \cite{tridiagonal,PRSA,horizon}.
Naturally, the details of such a realization remain nontrivial
even when we restrict our attention just to the
Wheeler-DeWitt form of the most elementary
differential-operator generators $G(t)$
and to the Big-Bang-like
quantum phase transitions.
Nevertheless, what we achieve is
that we avoid and eliminate
the danger of the
Big-Bounce smearing after quantization.
In Hermitian theory,
this smearing is unavoidable, caused by an effective level repulsion
as sampled in
Figure \ref{uba2} above.
In the quasi-Hermitian NIP context,
the Big-Bang-related exceptional-point degeneracy
is rendered possible via the
``fine-tuning'' of the metric: A few
non-numerical, exactly solvable simulations
of such a fine-tuning may be found described, e.g., in \cite{PRSA}.

A complementary word of warning has been formulated in our brief
methodical note \cite{Heisenberg}.
We revealed there that in Heisenberg picture (HP),
the Big-Bang degeneracy cannot be realized at all.
Indeed,
the underlying constat choice of vanishing $G^{(HP)}(t)=0$
has been shown to imply the stationarity of
the HP metric, $\Theta^{(HP)}\neq \Theta^{(HP)}(t)$
(recall Eq.~(\ref{2.6}) for the quick proof).
The HP form of
Eq.~(\ref{2.3}) implies that we have
$\Sigma_{(HP)}(t)=H_{(HP)}(t)$ so that
only the solution of the
Heisenberg Eqs.~(\ref{beda}) is needed.
The only advantage
of using the HP simplification
is that
both of the underlying
Schr\"{o}dinger equations
remain trivial.
Nevertheless, as long as the realization of
the Big Bang degeneracy necessarily requires that
the Hilbert-space metric $\Theta(t)$  has to vary with time,
the use of the full-fledged NIP formalism
with nontrivial $G^{(NIP)}(t)$
is unavoidable.

Unfortunately, no help has been reached
in an extended
Heisenberg picture (EHP, \cite{IJTP}).
In a slightly amended  formalism we proposed the use of
a constant-operator choice  of a {\em non-vanishing\,}
generator $G_{(EHP)}(t)=G_{(EHP)}(0)\neq 0$.
We found that the EHP formalism can already describe
the evolution equivalent to the
one generated by the manifestly time-dependent self-adjoint
quantum Hamiltonians $\mathfrak{h}(t)$
(cf. Abstract of Ref.~\cite{IJTP} or a rediscovery of this
possibility in \cite{FrFrith}). Still,
the description of the phase transitions (like Big Bang)
remained beyond the capacity of the amended EHP approach.
The full-fledged NIP is needed.

\subsubsection{The detailed WDW NIP recipe}

In \cite{NIP} we outlined some of the
details of the constructive treatment of the quantum phase transitions.
We pointed out that our ``dynamical input'' knowledge of
the non-observable Hamiltonian $G(t)$
enables us to solve
the pair of our
Schr\"{o}dinger Eqs.~(\ref{SEFip}) and~(\ref{d2.3})
at any initial conditions.
In this sense, every
initial $N-$plet
 \be
  |\psi_1(0)\kt\,,|\psi_2(0)\kt\,,\ldots\,,|\psi_N(0)\kt\,
  \label{inijedna}
 \ee
and
 \be
  |\psi_{1,\Theta}(0)\kt\,,
  |\psi_{2,\Theta}(0)\kt\,,\ldots\,,|\psi_{N,\Theta}(0)\kt\,
  \label{inidva}
 \ee
chosen at $t=0$ can be used to construct
the time-dependent $N-$plets of the kets
 \ben
  |\psi_1(t)\kt\,,|\psi_2(t)\kt\,,\ldots\,,|\psi_N(t)\kt\,
 \een
and
 \ben
  |\psi_{1,\Theta}(t)\kt\,,|\psi_{2,\Theta}(t)\kt\,,\ldots\,,
  |\psi_{N,\Theta}(t)\kt\,.
 \een
Under an elementary working
hypothesis of a finite-, $N-$dimensional Hilbert space,
the additional initial bi-orthonormality assumption
 \be
  \br \psi_{m,\Theta}(0)\, |\psi_{n}(0) \kt = \delta_{m,n}\,,\
  \ m,n = 1, 2, \ldots, N\,
 \label{complean0}
 \ee
and the completeness
 \be
 \sum_{n=1}^N |\psi_{n}(0)\kt \br \psi_{n,\Theta}(0) | = I\,
 \label{comple0}
 \ee
become immediately extended to all times $t>0$,
 \be
 \sum_{n=1}^N |\psi_{n}(t)\kt \br \psi_{n,\Theta}(t) | = I\,,
 \ \ \ \ \ \
 \br \psi_{1,\Theta}(t)\, |\psi_{n}(t) \kt = \delta_{m,n}\,,\
  \ m,n = 1, 2, \ldots, N\,.
 \label{comple}
 \ee
Also the time-dependent metric operator
$\Theta(t)$ acquires the standard representation in ${\cal
H}^{(F)}$,
 \be
 \Theta(t)=\sum_{n=1}^N
 |\psi_{n,\Theta}(t)\kt \,
  \br \psi_{n,\Theta}(t)
 | \,.
 \label{defifi}
 \ee
This means that the choice of a suitable
generator $G(t)$ and of the two suitable initial vector sets
(\ref{inijedna}) and (\ref{inidva}) with properties
(\ref{complean0}) and (\ref{comple0})
does not leave too much space for the further requirements concerning
the dynamics.

Fortunately, we come to the conclusion that
the space left by the NIP formalism is still sufficient
for our present purposes.
Indeed, in our construction
we started from the
assumption of the knowledge of a
preselected WDW
form of the generator $G(t)$.
Such a
specific ``kinematical-like'' input
information
is still not in conflict with the
Big-Bang dynamics.
Indeed,
such a version of the general NIP formalism
still admits the
use of the
formal
spectral representation of the observables.
In this sense, there
exist the two most important operators of our present interest.
The first one is the ``dynamical'',
observable-energy-representing  operator $H(t)$ called Hamiltonian.
In its
spectral representation of the form
 \be
 H(t) = \sum_{n=1}^N\,|\psi_{n}(t)\kt E_n(t) \br \psi_{n,\Theta}(t)
 |\,
 \label{lefdi}
 \ee
the choice of the energy eigenvalues $ E_n(t)$
remains unrestricted.

In a climax of the story,
an entirely analogous expansion should be finally introduced
in order to define
the complementary, ``kinematical'', background-representing
operator of a suitable ``geometry''
or ``spatial grid'' operator (\ref{redadag}). In its analogous
spectral representation
 \be
 Q(t) = \sum_{n=1}^N\,|\psi_{n}(t)\kt q_n(t) \br \psi_{n,\Theta}(t)
 |\,
 \label{sdlefdi}
 \ee
we are free to require
the validity of the
Big-Bang constraint (\ref{leaq}) imposed upon all of its
spatial background representing eigenvalues $q_n(t)$.

\section{Discussion\label{discussion}}

The non-Hermitian innovation
of the NSP framework
opened, in \cite{aliWDW}, the
way towards
a deeper understanding of the KG- and WDW-like quantum systems in stationary
approximation.
Later, the birth of the
more sophisticated non-Hermitian version of the Dirac's
interaction picture \cite{timedep}
seemed to be, initially, just an artificial mathematical exercise.
Nobody seemed to be willing to admit that
the NIP formalism might find an application
in quantum gravity.
The main reason was that in the most
advanced version of quantum gravity
(i.e., in the canonical LQG approach),
virtually all the results
seemed to indicate that
the classical Big Bang
singularity has to be replaced by its quantized Big Bounce
alternative.

Even among mathematicians it has been firmly believed that
the quantization must necessarily smear out the singularities
of the classical Einstein's general relativity \cite{Thiemann}.
In this sense,
before any return to the quantum Big Bang hypothesis
it was necessary to wait for
a renewal of its support in the
realistic LQG context \cite{nobounce}.
Naturally, the problem is technically complicated.
In this sense also the present
methodical support of the latter hypothesis
is just schematic and incomplete. In its framework
we only had to leave
many important phenomenological requirements
aside.
Let us now mention some of them in the form of brief comments.

\subsection{The background-independence requirement}

In the Isham's foreword preceding the Thiemann's comprehensive 2007
monograph on canonical
quantum gravity \cite{Thiemann} the Hamiltonian constraint
$\hat{H}\psi= 0$ {\it alias} ``the famous
Wheeler-DeWitt equation'' is
characterized as ``arguably one of the most elegant equations in
theoretical physics, and certainly one of the most mathematically
ill-defined.'' In the introductory part of the book itself
one reads, indeed, that
the sufficiently rigorous specification of
a suitable Hilbert space in which the
Wheeler-DeWitt operator
$\hat{H}$ would be defined represents one
of the most important unresolved theoretical
challenges.

The latter Hilbert-space problem may be found
thoroughly discussed in section 9.2 of
the Mostafazadeh's 2010 study \cite{ali}.
Even the authors of the LQG results admit that
such an
approach does not yet provide a fully consistent description of
the physical reality. Still, their
approach addresses, successfully,
the necessary background independence
of the theory \cite{Rovelli}. In some sense, such a requirement
should be incorporated
in any theory which pretends to be ``fundamental''
rather than just ``effective''.

From the point of our present approach based on the drastically
simplified WDW equation the
constructions which would be background-independent
were found feasible. In some sense,
such a requirement can be perceived as lying in the very center of the
NIP approach in which, admittedly, one starts from the knowledge of the
explicit WDW form of the operator $G(t)$, but in which the
theory admitsthe introduction of an
``observable background''.
Although our present spectral-representation definition
(\ref{sdlefdi}) of such  an {\em independent\,}
kinematical background may look rather abstract,
a more specific example may be sought, say,
in \cite{Batal} where a nontrivial coordinate/background
has constructively been obtained in an elementary dynamical model.

In our considerations
the role of a geometric background has been played
by the ``dynamical input'' operator $Q(t)$ sampled by a matrix in Eq.~(\ref{tyomed})
with the spectrum $q_n(t)$ simulating the ``observable'' spatial grid points and
guaranteeing the existence of the Big Bang singularity at $t=0$
(cf. Eq.~(\ref{leaq}) or Figure \ref{uba3}).
In discussion, one only
has to emphasize the mathematical subtlety of the
correspondence between the hidden Hermiticity of $Q(t)$
and the fine-tuned nature of the corresponding Hilbert-space metric $\Theta(t)$
which guarantees the unitarity of the system (i.e., of the
evolution of the Universe from the very beginning of its
observability and existence).

In the latter considerations the truly drastic simplifications
of the picture seem still absolutely necessary at present. Skipping, typically,
the Lorentz-covariance requirements and working with the models in which
the time is a parameter and in which, for methodical reasons, the ``Universe''
is one-dimensional and
discretized via a finite mesh of the time-dependent
grid points $q_j(t)$, $j=1,2,\ldots,N$. In such a  ``Universe''
just the degeneracy $q_j(t) \to 0$ in the classical-physics
Big Bang limit
$t \to 0^-$ is asked for.

\subsection{Problems with terminology}

The conventional belief that the avoided crossings of the eigenvalues
are generic is equivalent to the (usually, just tacit) assumption of
the time-independence of the physical inner-product metric $\Theta$.
In opposite direction, once we replace the
conventional self-adjoint grid-matrix  $\mathfrak{q}(t)$
by its isospectral but merely hiddenly Hermitian
partner
${Q(t)}$,
we discover the existence of a new freedom in the formalism
as carried by the Dyson map $\Omega(t)$.
As a consequence,
the existence of the singular Big Bang grid-point limit (\ref{leaq})
is rendered possible.

In the language of mathematics
the innovation lies in the enhancement of the
flexibility
of the dynamical laws.
One arrives at
the less usual, non-Hermitian NIP formulation
of quantum mechanics.
In its framework the unitary and closed quantum systems
may be defined via structures
using more than one inner product,
i.e., strictly speaking, more than one Hilbert space.

One of the most welcome consequences is an
enhancement of the flexibility while on dark side
one may find terminological misunderstandings.
In the phenomenologically oriented literature
several nicknames
denote more or less the same theory.
Thus, in different papers one encounters, e.g., the reference to the
quasi-Hermitian quantum theory \cite{Geyer,Dieudonne}, to the
pseudo-Hermitian quantum theory \cite{ali}, to the
non-Hermitian but ${\cal PT}-$symmetric quantum theory
(usually also with ${\cal P}$ = parity and
with ${\cal T}$ = time reversal \cite{Carl}), to the
three-Hilbert-space quantum theory \cite{SIGMA} or to the
crypto-Hermitian quantum theory \cite{Smilga},
etc.

\subsection{The danger of an over-determination of the dynamical input}

All of the latter approaches lead to a perceivable
gain in flexibility of the realistic models of various quantum systems.
This is to be
paid merely by the necessity of keeping trace of
the more sophisticated forms of
Hermitian conjugations.
One can conclude that the subject is still hot.
On a model-independent level of discussion
it is worth adding that  the consistence of the dynamical input
need not in fact be easily guaranteed.
In review  \cite{Geyer}, for example, the authors
stressed that in the over-determined cases
the necessary Hilbert space metric
(and, hence, the theory itself)
need not exist at all.
In \cite{arabky}, such a non-existence has been shown to occur even in some
fairly popular realistic models. An abstract analysis of such an
unpleasant possibility was presented in \cite{Lotoreichik}.
Only recently, more encouraging results were obtained in \cite{EPJP}
offering a certain systematic guide to the construction
of the mutually compatible non-Hermitian observables.

Once we restrict attention to the applicability of the NIP approach in
cosmology, an encouragement may be sought in the
progress and
simplifications of the canonical quantization
\cite{Thiemann,Rovelli}.
The latter two reviews of the state of art
differ by the language, with the former one being more
mathematically oriented. Still, both of these monographs
share the traditional philosophy interpreting the
quantum theory as a result of a modification of its classical
predecessor.
In our final remark we would like to
point out that
one could also try to weaken our dependence
on the  classical-physics-based intuition
by treating, as primary, the tentative quantum hypotheses
in a way
defended, e.g., by Brody and Hughston \cite{Dorje}.

\section{Summary  \label{seci5}}

The core of our present message is that
the consistency of the quantum-mechanical interpretation of the
non-stationary WDW systems requires that
Schr\"{o}dinger equation ceases to
be perceived as offering a complete picture
of the evolution. In this sense, it is necessary to add
a parallel and full-fledged description of the evolution of the operators
of observables using the Heisenberg-like evolution
equations.
In the natural physical quantum-gravity context,
the unitarity of the WDW-controlled
evolution can be then guaranteed. The
apparently non-unitary evolution of the left and right
wavefunctions
(controlled by the
respective two Schr\"{o}dinger-type equation) is
precisely compensated by the apparent non-unitarity of
the evolution of the operators
representing the observables (controlled, in parallel, by non-Hermitian
Heisenberg-type equations).

Having accepted such a philosophy,
our present paper can be read as
a more or less purely methodical
return to the question whether, in
the framework of quantum cosmology, the
birth of our Universe should be perceived as
a point-like Big Bang or as a smeared Big Bounce.
In essence, we presented here
a few arguments supporting our persuasion that in the
purely theoretical NIP framework such a question remains,
at present, open.

\section*{Appendix: Two Hilbert spaces in
 quantum mechanics.}

In the conventional quantum mechanics of textbooks \cite{Messiah}
the predictions of the results of experiments have their
mathematical background, in NSP, in the evaluation of
matrix elements
 \be
 \pbr \psi^{(NSP)}(t_f)\,|\mathfrak{q}_{(NSP)}(t_f)\, |\psi^{(NSP)}(t_f)
 \pkt\,.
 \label{MEAS}
 \ee
The symbol
$\mathfrak{q}_{(NSP)}$
denotes here a self-adjoint operator
of the
observable in question: Usually, this operator is time-independent,
$\mathfrak{q}_{(NSP)} \neq \mathfrak{q}_{(NSP)}(t)$.
All information about the evolution of the system in time
is carried, in the pure state regime, by a ket-vector element
 $|\psi^{(NSP)}(t)\pkt$
of a Hilbert space of states ${\cal H}^{(textbook)}$.
This state is assumed prepared at $t_{initial}=0$
and measured at $t=t_{final}=t_f$.
Prediction (\ref{MEAS}) is
probabilistic and contains
just the NSP wave-ket solutions
$|\psi^{(NSP)}(t)\pkt$ of Schr\"{o}dinger equation
 \be
 {\rm i} \frac{\partial}{\partial t} \,|\psi(t)\pkt =
 \mathfrak{h}_{(NSP)}\,|\psi(t)\pkt\,,
 \ \ \ \ \
 \ \ \ \ \ |\psi(t)\pkt \in {\cal H}^{(textbook)}\,.
 \label{SET}
 \ee
Due to the Stone
theorem
the evolution is unitary if and only if
the Hamiltonian is self-adjoint in
${\cal H}^{(textbook)}$,
$\mathfrak{h}_{(NSP)}=\mathfrak{h}_{(NSP)}^\dagger\,$ \cite{Stone}.

One of many
efficient simplifications of the practical solution of
Eq.~(\ref{SET}) is due to Dyson \cite{Dyson}.
He revealed that in many cases
one has to work with a technically unfriendly
Hamiltonian which can be perceivably simplified
via a suitable isospectral preconditioning
$\mathfrak{h}_{(NSP)} \to  H_{(NSP)} =\Omega^{-1}\, \mathfrak{h}_{(NSP)}\,\Omega_{}$.
This is formally equivalent to the
transformation of the ket-vector wave functions,
 \be
 |\psi_n^{(textbook)}\pkt=\Omega_{}\,|\psi_n^{(auxiliary)}\kt\,,
 \ \ \ \ n=0,1,\ldots
 \,.
 \label{lumapo}
 \ee
Operator $\Omega_{}$
has to be $\,n-$independent and stationary ($\Omega_{} \neq
\Omega_{}(t)$).
Dyson also recommended to make the choice of $\Omega_{}$
non-unitary
($\Omega^\dagger_{}\Omega_{}=\Theta_{} \neq I$).
In analogy with the so called coupled-cluster method
based on a similar idea \cite{Bishop},
one may
also treat
the simpler partner of the
Hilbert space ${\cal H}^{(textbook)}$
as formally different, denoted
by a different dedicated symbol, say,
${\cal H}^{(friendlier)}$.

Schr\"{o}dinger
Eq.~(\ref{SET}) becomes replaced,
in the majority of applications of the Dyson-recommended
and $\Omega-$mediated
change of space
${\cal H}^{(textbook)} \to {\cal H}^{(friendlier)}$, by a
friendlier equation
 \be
 {\rm i} \frac{\partial}{\partial t} \,|\psi^{(Dyson)}(t)\kt
 =H_{(Dyson)}\,|\psi^{(Dyson)}(t)\kt\,,
 \ \ \ \ \ \
 |\psi^{(Dyson)}(t)\kt \in {\cal H}^{(friendlier)}\,.
 \label{SETdys}
 \ee
The
transformed Hamiltonian is
de-Hermitized since
$H = \Omega^{-1}\,\mathfrak{h}\,\Omega
\neq H^\dagger$ in ${\cal H}^{(friendlier)}$.
In the early review \cite{Geyer} of the procedure
a change of the philosophy has been proposed,
resulting in a reformulation of the textbook NSP approach
called,
in the spirit of the mathematician's terminology \cite{Dieudonne},
quasi-Hermitian quantum mechanics.
In this framework
the model-building process has to
start directly
from Eq.~(\ref{SETdys})
and from a guarantee of the user-friendliness of
the preconditioned Hamiltonian
$H$. Whenever necessary, one may, after all, re-Hermitize the model, say,
via a reconstruction of $\Omega$ \cite{ali}.

The non-unitarity of the map $\Omega$
implies,
for the manifestly auxiliary Hilbert space ${\cal H}^{(friendlier)}$,
the loss of its physical-space status. Fortunately,
it appeared sufficient to amend the inner product
and to convert ${\cal H}^{(friendlier)}$ into a fully
acceptable and physical Hilbert space
${\cal H}^{(standard)}$.
By construction, the latter space has to be unitary
equivalent to ${\cal H}^{(textbook)}$, with the most straightforward
method being
the reconstruction
of the so called metric $\Theta=\Omega^\dagger\,\Omega$. The
mathematical details can be found in reviews \cite{Geyer} or \cite{ali}.
The essence of the trick is that
the correct space ${\cal H}^{(standard)}$
can be represented
via the mere amendment of the bra vectors in  ${\cal H}^{(friendlier)}$,
 \be
 \br \psi | \ \to \  \br \psi | \Theta_{}
  \ \equiv \ \br \psi_\Theta |
 \ \ \ \ \ {\rm for} \ \ \ \ \
 {\cal H}^{(friendlier)} \to {\cal H}^{(standard)}
 \,.
 \label{amelio}
 \ee
In the terminology of
functional analysis the definition of the dual {\it
alias} bra-vector space of the linear functionals is
merely amended and transferred back, from  ${\cal H}^{(standard)}$
to ${\cal H}^{(friendlier)}$,
via formula ${\cal V}' \to
{\cal V}'_\Theta\,$. In other words,
one just converts the conventional, unphysical
bra-ket inner product $\br \psi |\chi \kt$
into its physical alternative,
 \be
 \br \psi
 |\chi \kt \to \br \psi|\Theta |\chi \kt \ \equiv \ \br \psi_\Theta
 |\chi \kt\,.
  \ee
In the light of this relation
it is possible to perform all calculations
in ${\cal H}^{(friendlier)}$.
Still, in practice,
the redundancy of the introduction of
the manifestly unphysical Hilbert space ${\cal H}^{(friendlier)}$
must be well motivated. The expense
must  be
more than compensated by the simplification of the
evaluation of the experimental predictions.
Also the loss of the direct connection
with ${\cal
H}^{(textbook)}$ has to be taken into account
because in this space
we usually
define the operators of observables using
the principle of correspondence \cite{Messiah}.

One can often pull
at least some of the necessary operators from ${\cal
H}^{(textbook)}$
up to the
auxiliary Hilbert space ${\cal H}^{(friendlier)}$
(see, e.g., \cite{Batal}). E.g., whenever one knows the
Dyson map,
one can define the necessary operators
in ${\cal
H}^{(friendlier)}$ using formula
 \be
 Q_{(Dyson)}= \Omega_{(Dyson)}^{(-1)}\,
 \mathfrak{q}_{(NSP)}\,\Omega_{(Dyson)} \neq Q_{(Dyson)}^\dagger\,.
 \label{dadag}
 \ee
The experiment-predicting NSP formulae
(\ref{MEAS}) then acquire the upgraded forms,
 \be
 \pbr \psi^{(NSP)}(t_f)\,|\mathfrak{q}_{(NSP)}(t_f)\, |\psi^{(NSP)}(t_f)
 \pkt=
 \br \psi^{(Dyson)}_\Theta(t_f)\,|Q_{(Dyson)}(t_f)\,
 |\psi^{(Dyson)}(t_f)
 \kt\,
 \label{dysMEAS}
  \ee
in which
one can use, at worst, just some reasonably precise
approximate forms
of the physical Hilbert space metric $\Theta=\Omega^\dagger\,\Omega$
in Eq.~(\ref{amelio})
(cf. \cite{Geyer,Batalb}).

.

 \noindent
{\bf Funding: }
{{This work is supported by the  Faculty of Science of the
University of Hradec Kr\'{a}lov\'{e} and by the
Nuclear Physics
Institute in \v{R}e\v{z}.}}

.

 \noindent
{\bf Conflicts of Interest: }
{{The author declares no conflict of interest.}}


\newpage

\begin{thebibliography}{99}


\bibitem{WDW}
 DeWitt, B. S. Quantum Theory of Gravity.
 I. The Canonical Theory.
\emph{Phys. Rev.}
 {\bf 1967},
 {\em 160},
     1113--1148.

\bibitem{WDWb}
Hamber, H. W.;  Williams, R. M.
Discrete Wheeler-DeWitt Equation.
\emph{Phys. Rev.}
 {\bf 2011},
 {\em 84},
     104033.

\bibitem{Thiemann}
Thiemann, T.
\emph{Introduction to Modern Canonical Quantum General Relativity};
Cambridge University Press: Cambridge, UK, 2007.

\bibitem{Rovellib}
Rovelli, C.;  Smolin, L.
  Loop space representation of
quantum general relativity.
\emph{Nucl. Phys. B}
 {\bf 1990},
 {\em 331},
     80--152.

\bibitem{Rovelli}
Rovelli, C. \emph{Quantum Gravity}; Cambridge University Press: Cambridge, UK, 2004.



\bibitem{Rovellic}
Ashtekar, A.; Lewandowski, J.
 Background independent
quantum gravity: a status report.
\emph{Class. Quantum Grav.}
 {\bf 2004},
 {\em 21},
     R53--R152.




\bibitem{ali}
Mostafazadeh, A. Pseudo-Hermitian Representation of
Quantum Mechanics.
{\emph{Int. J. Geom. Meth. Mod. Phys.}} \textbf{2010}, \emph{7}, 1191--1306.



\bibitem{aliWDW}
%
Mostafazadeh, A.
Quantum mechanics of Klein-Gordon-type fields and quantum cosmology.
\emph{Ann. Phys. (N.Y.)}
 {\bf 2004},
 {\em 309},
     1--48.

%
\bibitem{which}
%
%
%
%
%
M. Znojil, \emph{
Which operator generates time evolution in Quantum
Theory?}
arXiv:
quant-ph 0711.0535.


\bibitem{timedep}
Znojil, M.
Time-dependent version of cryptohermitian quantum theory.
%
%
%
\emph{Phys. Rev. D}
 {\bf 2008},
 {\em 78},
     085003.


\bibitem{SIGMA}
{Znojil, M. Three-Hilbert-space} formulation of Quantum Mechanics.
\emph{Symm. Integ. Geom. Meth. Appl. SIGMA} {\bf 2009}, {\em 5}, 001.



\bibitem{NIP}
Znojil, M. Non-Hermitian interaction representation and its use in relativistic quantum mechanics.
\emph{Ann. Phys. (NY)} {\bf 2017}, {\em 385}, 162--179.


\bibitem{AshteBi}
Bojowald, M.
Absence of a singularity in loop quantum cosmology.
\emph{Phys. Rev. Lett.}
 {\bf 2001},
 {\em 86},
    5227--5230.

%
%
%
%
%
%
%


\bibitem{loopash}
 Ashtekar, A.;  Pawlowski, T.;  Singh, P. Quantum nature of the big
bang: Improved dynamics. \emph{Phys. Rev. D} \textbf{2006},
\emph{74},   084003.


\bibitem{loopashb}
Bojowald, M.
Quantum nature of cosmological bounces.
\emph{Gen. Rel. Grav.}
 {\bf 2008},
 {\em 40},
     2659--2683.


\bibitem{loopashc}
 Ashtekar,  A.;  Corichi, A.;  Singh, P. Robustness of key features of
loop quantum cosmology. \emph{Phys. Rev. D}  \textbf{2008},
\emph{77}, 024046.



%
\bibitem{piech}
 Malkiewicz, P.;  Piechocki, W. Turning Big Bang into Big Bounce:
 II. Quantum dynamics. \emph{Class. Quant. Gravity} \textbf{2010}, \emph{27},  225018.

%
%


\bibitem{loopashd}
Bojowald, M.; Paily, G. M.
A no-singularity scenario in loop quantum gravity.
\emph{Class. Quant. Gravity}
 {\bf 2012},
 {\em 29},
     242002.





\bibitem{loopashe}
Yang, J.-S.; Zhang, C.; Ma, Y.-G.
Loop quantum cosmology from an alternative Hamiltonian.
\emph{Phys. Rev. D}
 {\bf 2019},
 {\em 100},
     064026.




\bibitem{nobounce}
 Wang, Ch.; Stankiewicz, M.
  Quantization of time and the big bang
  via scale-invariant loop gravity.
\emph{Phys. Lett. B}
 {\bf 2020},
 {\em 800},
     135106.


\bibitem{Messiah}
Messiah, A. \emph{Quantum Mechanics};
North Holland: Amsterdam, The Netherlands, 1961.




\bibitem{Geyer}
%
Scholtz, F. G.;  Geyer, H. B.;
Hahne,  F. J. W.
Quasi-Hermitian Operators in Quantum Mechanics and the Variational Principle.
\emph{Ann. Phys. (NY)} {\bf 1992}, {\em 213}, 74--101.
%


\bibitem{book}
%
Bagarello, F.; Gazeau, J.-P.; Szafraniec, F.; Znojil, M. (Eds.)
\emph{Non-Selfadjoint Operators in Quantum Physics: Mathematical Aspects};
Wiley: {Hoboken, NJ, USA,} 
2015.


\bibitem{Carlbook}
Bender, C.M. \emph{PT Symmetry in Quantum and Classical Physics};
World Scientific: Singapore, 2018.


\bibitem{EPJP}
Znojil. M.
Feasibility and method of multi-step Hermitization of crypto-Hermitian quantum
Hamiltonians.
\emph{Eur. Phys. J. Plus}
 {\bf 2022},
 {\em 137},
     335.

\bibitem{grid}
Rovelli, C.;  Smolin, L.  Discreteness of area and volume
in quantum gravity.
\emph{Nucl. Phys. B}
 {\bf 1995},
 {\em 442},
     593--619.

\bibitem{gridb}
Thiemann, T.  A length operator for canonical quantum
gravity.
\emph{J. Math. Phys.}
 {\bf 1998},
 {\em 39},
     3372--3392.


\bibitem{gridc}
 Znojil, M. Quantum Big Bang without fine-tuning in a toy-model. \emph{J.
Phys. Conf. Ser.}  \textbf{2012}, {\em 343}, 012136.



\bibitem{gridd}
Znojil, M.
Quantization of Big Bang in crypto-Hermitian Heisenberg
picture,  in F. Bagarello, R. Passante and C. Trapani, eds,
\emph{Non-Hermitian Hamiltonians in Quantum Physics.
}; Springer: Cham,
2016;
series
\emph{Springer Proceedings in Physics}
 {\bf 2016},
 {\em 184},
     383--399.




\bibitem{gride}
Brody, D. C.; Hughston, L. P.
Quantum measurement of space-time events.
\emph{J. Phys. A: Math. Theor.}
 {\bf 2021},
 {\em 54},
     235304.

\bibitem{Kato}
Kato, T.
\emph{Perturbation Theory for Linear Operators};
Springer: Berlin/Heidelberg, Germany,
 1966.


\bibitem{symmetry}
Znojil, M.
Parity-time symmetry and the toy models of gain-loss
dynamics near the real Kato's exceptional points.
\emph{Symmetry}
 {\bf 2016},
 {\em 8},
     52.


\bibitem{nine}
Styer, D. F.; et al.
Nine formulations of quantum mechanics.
\emph{Am. J. Phys.}
 {\bf 2002},
 {\em 70},
    288 -- 297.

\bibitem{Nimrod}
Moiseyev, N. \emph{Non-Hermitian Quantum Mechanics};
Cambridge University Press: Cambridge, UK, 2011.



















\bibitem{FV}
Feshbach, H.;  Villars, F.
Elementary relativistic wave
mechanics of spin 0 and spin 1/2 particles.
%
%
\emph{Rev. Mod. Phys.}
 {\bf 1958},
 {\em 30},
     24--45.


\bibitem{WDWja}
Znojil, M.
Relativistic supersymmetric quantum mechanics based on Klein-Gordon equation.
%
%
%
%
\emph{J. Phys. A: Math. Gen.}
 {\bf 2004},
 {\em 37},
     9557--9571.

\bibitem{PW}
Pauli, W.;  Weisskopf, V.
Uber die Quantisierung der skalaren relativistischen Wellengleichung.
\emph{Helv. Phys. Acta}
 {\bf 1934},
 {\em 7},
     709--731.


\bibitem{Carl}
Bender, C. M. Making sense of non-Hermitian Hamiltonians. \emph{Rep.
Prog. Phys.} {\bf 2007}, {\em 70}, 947--1118.


\bibitem{[150]}
Mostafazadeh, A.
Hilbert space structures on the solution space of Klein-Gordon type evolution equations.
\emph{Class. Quant. Grav.}
 {\bf 2003},
 {\em 20},
     155--171.








\bibitem{AshtekarBi}
Gielen, S.; Turok, N.
Perfect Quantum Cosmological Bounce.
\emph{Phys. Rev. Lett.}
 {\bf 2016},
 {\em 117},
     021301.




\bibitem{AshtekarBib}
Ashtekar, A.;   Bianchi, E.
A short review of loop quantum gravity.
\emph{Rep. Prog. Phys.}
 {\bf 2021},
 {\em 84},
     042001.
%
%



\bibitem{recurrently}
M. Znojil,
Quantum inner-product metrics via recurrent solution of Dieudonne equation.
 J. Phys. A: Math. Theor. 45, 085302 (2012);


\bibitem{tridiagonal}
 Znojil, M. Tridiagonal PT-symmetric \emph{N} by \emph{N} Hamiltonians and a
fine-tuning of their observability domains in the strongly
non-Hermitian regime.
%
\emph{J. Phys. A Math. Theor.} \textbf{2007}, \emph{40},
13131--13148.
%
%


\bibitem{minimal}
 Znojil, M. Maximal couplings in PT-symmetric chain-models
with the real spectrum of energies.
%
\emph{J. Phys. A Math. Theor.} \textbf{2007}, \emph{40},
4863 --
4875.
%
%


\bibitem{annalsix3}
Znojil, M. N-site-lattice analogues of $V (x) = ix^3$.
\emph{Ann. Phys. (N.Y.)}
 {\bf 2012},
 {\em 327},
    893--913.


\bibitem{SIGMAdva}
Znojil, M. On the role of the normalization factors $\kappa_n$ and of
the pseudo-metric $P$ in crypto-Hermitian quantum models.
\emph{Symm. Integ. Geom. Meth. Appl. SIGMA} {\bf 2008}, {\em 4}, 001.


\bibitem{Heisenberg}
Znojil, M.
Non-Hermitian Heisenberg representation.
\emph{Phys. Lett. A}
 {\bf 2015},
 {\em 379},
    2013--2017.

\bibitem{cinani}
Miao, Y.-G.; Xu, Zh.-M.
Investigation of non-Hermitian Hamiltonians in the Heisenberg Picture.
\emph{ Phys. Lett. A}
 {\bf 2016},
 {\em 380},
     1805--1810.

\bibitem{FringMou}
Fring, A.;  Moussa, M. H. Y.
%
%
Unitary quantum evolution for time-dependent quasi-Hermitian systems
%
with non-observable Hamiltonians.
\emph{Phys. Rev. A}
 {\bf 2016},
 {\em 93},
     042114.


%
%

%



\bibitem{Luiz}
%
 Luiz,  F. S.; Pontes, M. A.; Moussa, M. H. Y.
Unitarity of the time-evolution and observability of non-Hermitian
Hamiltonians for time-dependent Dyson maps.
arXiv: 1611.08286.
%
%


\bibitem{Wang}
Gong, J.-B.;   Wang, Q.-H.
Time-dependent PT-symmetric quantum
mechanics.
 \emph{J. Phys. A: Math. Theor.}
 {\bf 2013},
 {\em 46},
    485302.




%




\bibitem{Bila}
B\'{\i}la, H.
 \emph{Non-Hermitian Operators in
Quantum Physics}: Charles University, Prague, 2008 (PhD thesis).


\bibitem{Bilab}
B\'{\i}la, H.
\emph{Adiabatic time-dependent metrics in PT-symmetric
quantum theories}:
  e-print
 arXiv: 0902.0474.



\bibitem{FrFrith}
Fring, A.; Frith, T.
Exact analytical solutions for time-dependent
Hermitian Hamiltonian systems from static unobservable non-Hermitian Hamiltonians.
\emph{Phys. Rev. A}
 {\bf 2017},
 {\em 95},
     010102(R).

\bibitem{IJTP}
Znojil, M.
Crypto-unitary forms of quantum evolution operators.
\emph{Int. J. Theor. Phys.}
 {\bf  2013},
 {\em 52},
     2038.

\bibitem{PRSA}
Znojil, M.
Passage through exceptional point: Case study.
\emph{Proc. Roy. Soc. A: Math. Phys.
Eng. Sci.}
 {\bf 2020},
 {\em 476},
      20190831.




\bibitem{horizon}
Znojil, M. Horizons of stability. \emph{J. Phys. A Math. Theor.}
\textbf{2008}, \emph{41}, 44027.

%





\bibitem{Batal}
Mostafazadeh,  A.; Batal,  A. Physical Aspects of Pseudo-Hermitian
and PT-Symmetric Quantum Mechanics. \emph{J. Phys. A: Math. Gen.}
 {\bf 2004},
 {\em 37},
    11645--11679.


\bibitem{Dieudonne}
Dieudonne, J. Quasi-Hermitian Operators. In
\emph{Proc. Int. Symp. Lin. Spaces}, Pergamon: Oxford, UK, 1961,
{pp. 115--122}.

\bibitem{Smilga}
Smilga, A.V. Cryptogauge symmetry and cryptoghosts for crypto-Hermitian Hamiltonians.
\emph{J. Phys. A Math. Theor.} {\bf 2008}, {\em 41}, 244026.



\bibitem{arabky}
Znojil, M.; Semor\'{a}dov\'{a}, I.; R\u{u}\v{z}i\v{c}ka, F.; Moulla, H; Leghrib, I.
Problem of the coexistence of several non-Hermitian observables in PT-symmetric quantum mechanics.
\emph{Phys. Rev. A} {\bf 2017}, {\em 95}, 042122.


\bibitem{Lotoreichik}
Krej\v{c}i\v{r}\'{\i}k, D.; Lotoreichik, V.; Znojil, M. The minimally anisotropic
metric operator in quasi-hermitian quantum mechanics.
\emph{Proc. Roy. Soc. A Math. Phys. Eng. Sci.} {\bf 2018}, {\em 474}, 20180264.


\bibitem{Dorje}
Brody, D. C.;  Hughston, L. P.
Geometric quantum mechanics.
\emph{J. Geom. Phys.}
 {\bf 2001},
 {\em 38},
     19--53.

\bibitem{Stone}
Stone, M.H. On one-parameter unitary groups in Hilbert Space.
\emph{Ann. Math.} {\bf 1932}, {\em 33}, 643--648.

\bibitem{Dyson}
%
%
Dyson, F.J. General Theory of Spin-Wave Interactions. \emph{Phys.
Rev.} {\bf 1956}, {\em 102}, 1217--1230.


\bibitem{Bishop}
Bishop, R. F.; Znojil, M. The coupled-cluster approach to quantum
many-body problem in a three-Hilbert-space reinterpretation.
\emph{Acta Polytech.} {\bf 2014}, {\em 54}, 85--92.



\bibitem{Batalb}
Znojil, M.
The cryptohermitian smeared-coordinate representation of wave
functions.
\emph{Phys. Lett. A}  \textbf{2011},  \emph{375}, 3176--3183.





\end{thebibliography}
 \end{document}